\documentclass[onecolumn,prd,amsfonts,epsfig,floats,superscriptaddress,floatfix,nofootinbib]{revtex4}
\usepackage{graphicx}
\usepackage{dcolumn}
\usepackage{bm}
\usepackage{longtable}
\usepackage{amsmath}
\usepackage{mathrsfs}
\usepackage{epstopdf}
\graphicspath{{./}}

\begin{document}
\bibliographystyle{apsrev}


\title{Mapping the galactic gravitational potential with peculiar acceleration}

\author{Claudia Quercellini} 
\affiliation{Dipartimento di Fisica, Universit\`a di Roma ``Tor Vergata'', via della Ricerca Scientifica 1, 00133 Roma, Italy}
\author{Luca Amendola}
\affiliation{INAF/Osservatorio Astronomico di Roma, Via Frascati 33, 00040 Monte
Porzio Catone, Roma, Italy}
\author{Amedeo Balbi}
\affiliation{Dipartimento di Fisica, Universit\`a di Roma ``Tor Vergata'', via della Ricerca Scientifica 1, 00133 Roma, Italy}
\affiliation{INFN, Sezione di Roma ``Tor Vergata'', via della Ricerca Scientifica 1, 00133 Roma, Italy}

\begin{abstract}
It has been suggested recently that the change in cosmological redshift (the Sandage test of expansion) could be observed in the next generation of large telescopes and ultra-stable spectrographs. In a recent paper we estimated the
change of peculiar velocity, i.e.\ the peculiar acceleration, in nearby galaxies and clusters and shown it to be of the same order of magnitude as the typical cosmological signal. Mapping the acceleration field allows for a reconstruction of the galactic gravitational potential without assuming virialization. In this paper we focus on the peculiar acceleration in our own Galaxy, modeled as a Kuzmin disc and a dark matter spherical halo. We estimate the peculiar acceleration for all known Galactic globular clusters and find some cases with an expected velocity shift  in excess of 20 cm/sec for  observations fifteen years apart, well above the typical cosmological acceleration. We then compare the predicted signal for a MOND (modified Newtonian dynamics)  model in which the spherical dark matter halo is absent. We find that the signal pattern is qualitatively different, showing that the peculiar acceleration field could be employed to test competing theories of gravity. However the difference seems too small to be detectable in the near future.
\end{abstract}

\maketitle

\section{Introduction}
\label{intro}
Present cosmology has plenty of accurate data on the cosmic expansion,
from standard candles to big bang thermal relics, but still lacks a direct observation of expansion which does not rely on interpreting the redshift as due the expansion itself. Such a direct test of expansion has been
suggested in the 60's by Sandage \citep{sandage62} and recently reproposed by various
authors \citep*{loeb98,pasquini06,2007PhRvD..76f3508L,2007MNRAS.382.1623B,2008PhRvL.100s1303U,2008PhRvD..77b1301U,2007astro.ph..1433C} in view of tremendous technological advances in spectroscopy and telescope collecting power.
The expansion can be directly observed by measuring the redshift of
sources at cosmological distances (e.g.\ quasars) at two different epochs
in time, obtaining for a homogeneous and isotropic Universe the relation
\begin{equation}
\dot{z}=(1+z)H_{0}-H(z)\end{equation}
The expected effect at $z\approx1$ amounts to less than a centimeter per second per year. Today's spectroscopy
has reached already a sensitivity of few meters per second: by
using spectra populated by many sharp lines, some authors have shown that a statistical sensitivity of the required level can be reached with the next generation of optical telescopes \citep{2008MNRAS.386.1192L}.

The total signal $\dot{z}$ results clearly from the combination of
the cosmological acceleration and of the line-of-sight component of
the local (peculiar) acceleration field, just as the redshift itself
combines global and local velocities. At large scales ($z\ge0.5$)
the peculiar acceleration signal acts as a noise over the cosmological signal and it might be hard to discriminate between the two contributions. At smaller depths however the peculiar field dominates and can be observed directly. In a recent paper \citep*{2008PhLB..660...81A} we found that the peculiar acceleration in nearby clusters and galaxies is in fact of the same order of magnitude of the cosmological signal at larger distances and could be measured with the same instrumentation.

In this paper we focus instead on the peculiar acceleration of our
own Galaxy and see how this can help reconstructing the Milky Way
gravitational potential. We model our Galaxy as a Kuzmin disk plus
a spherical dark halo and calculate the expected peculiar acceleration
as a function of source distance and Galactic coordinates. Then we
apply the general expression to all known Galactic globular clusters,
finding signals up to 20 cm/sec in 15 years, quite larger than the typical expected cosmological signal.

The peculiar acceleration field maps directly the gravitational potential
without the need to rely on virialization assumptions. One is therefore
led to ask whether the peculiar acceleration field could help in discriminating
among competing gravitational theories. A well-known model of non-Newtonian gravity is the modified Newtonian dynamics (MOND) theory \citep{Brada:1995}: in this theory, the galaxy can be
modeled as a baryonic disk, without any spherical halo, with parameters
chosen to obtain the same asymptotic disk rotational velocities. We
estimated the Milky Way peculiar acceleration field in MOND and compared
it to the Newtonian field for sources outside the disk, where the
geometry of the two models should maximize the difference. We find
qualitatively distinct patterns in the two cases but we find rather
small quantitative differences, not larger than 1 cm/sec, probably
too low to be within the reach of foreseeable observations. It is
of course possible that peculiar acceleration can be used to discriminate
Newtonian dynamics from other models of gravity in our own Galaxy,
thereby extending the test of gravity far beyond the present solar
system limits. Furthermore, measuring peculiar accelerations could be a way to determine the gravitational potential parameters.

\section[]{Peculiar acceleration}
\label{pec}

We start by investigating the peculiar acceleration signal expected in the case of a generic spiral galaxy. Later, in Section \ref{milkyway}, we will address the specific case of our own Galaxy. 

The signal we are after is the velocity shift $\Delta v$ given by
\begin{equation}
\Delta v=a_{s} \Delta T,
\end{equation}
where $a_s$ is the line of sight component of the acceleration and $\Delta T$ is the time interval between two observations of the velocity of a test particle.

Let us first consider the disc component of the galaxy. 
If a test particle (e.g.\ stars, gas) orbits on the disc, a mathematical theorem states that, due to the symmetry of the distribution, its peculiar acceleration is affected only by the mass embedded within its distance from the centre, and it is the same as if this mass was totally concentrated in the centre. 

In order to model the disc component, we will consider the so-called Kuzmin disc, namely a disc with superficial density
\begin{equation}
 \label{densi}
 \Sigma(r)=\frac{hM}{2\pi(R^{2}+h^{2})^{3/2}},
\end{equation}
where M is the total disc mass, $R$ and $z$ are cylindrical coordinates and $h$ is the scale length of the disc. 
The two-parameter Newtonian gravitational potential outside the disc is
\begin{equation}
 \label{kuzpot}
 \phi_{K}=-\frac{MG}{[R^{2}+(|z|+h)^{2}]^{1/2}},
\end{equation}
and the equipotential surfaces are concentric spheres centered at $\pm h$ (see Fig.~\ref{fig:ill1}). The test particle acceleration then reads
\begin{equation}
 \label{acc2}
 a_{K}=-\frac{MG}{[R^{2}+(|z|+h)^{2}]},
\end{equation}
with $|z|$ accounting for the acceleration field above and below the disc.  The force field of a Kuzmin disc no longer converges towards the origin of axis (see Fig.~\ref{fig:ill1}). Consequently, the projection angle $\theta'$ must be corrected by $\gamma$, namely the angle between the acceleration and the radial direction $r$.

On the other hand the peculiar acceleration of a test particle outside the disc, e.g.\ a globular cluster, is not only affected by the  gravitational potential generated by the disc, but also by the possible presence of a dark matter halo needed to explain the observed rotation curves. An alternative description of the gravitational potential can be given within the MOND framework: in this case, only baryonic matter in the disc is present, and the gravitational law is modified. The peculiar acceleration of a test particle outside the disc will then be affected by the axisymmetrical gravitational potential of the disc. 

We will now calculate the expected signal in the two scenarios: first, the standard configuration of a disc plus a cold dark matter halo; second, the alternative description in terms of a disc plus MOND. The bulge component will be neglected in both scenarios, since its spherical symmetry allows one to treat it simply as an additional contribution to the total mass for scales outside the bulge.

\subsection{Disc + CDM halo}
 \label{CDM}
In standard Newtonian mechanics we model the galaxy as
a Kuzmin disc embedded in a CDM halo, described by a spherical logarithmic potential, following  \cite{Read:2005}. Thus we adopt the Kuzmin potential for the disc (as in Eq.~\ref{kuzpot}) and we model the halo potential as logarithmic,
\begin{equation}
\label{halo}
\phi_{L}=\frac{1}{2}v_{0}^{2}\log{\Big(R_{c}^{2}+R^{2}+\frac{z^{2}}{q}\Big)},
 \end{equation}
where $R_{c}$ is the scalelength, $q$ is the halo  flattening ($q=1$ recovers  spherical symmetry) and $v_{0}$ is the asymptotic value of the velocity at large radii. 

\begin{figure}
  \centering
    \includegraphics[width=8.truecm]{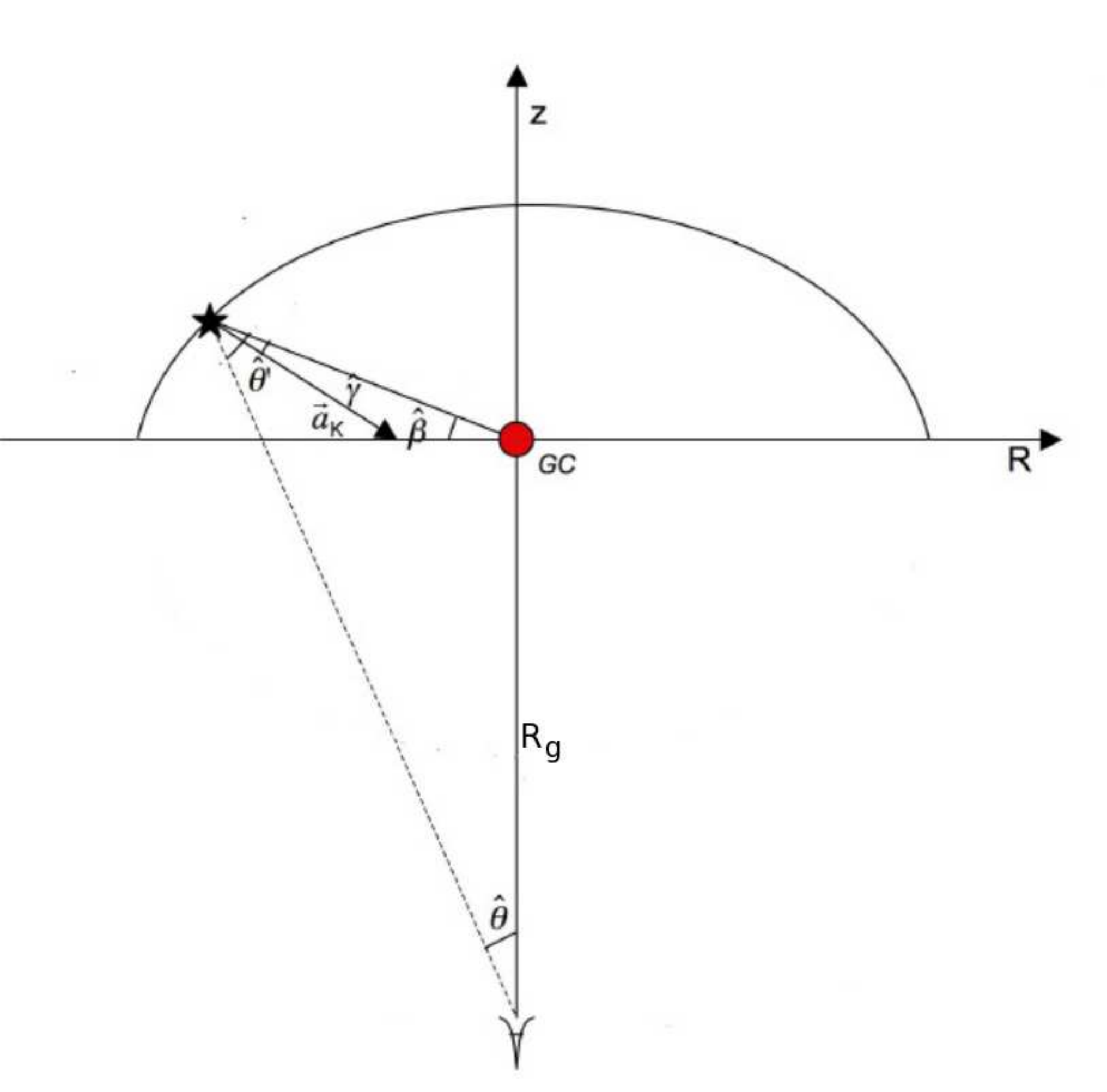}
    \caption{ \small \small Kuzmin disc potential in cylindrical coordinates. The disc lies edge-on on the axis $R$. The contour shows an equipotential curve and $\vec{a}_{K}$ represents the  force  field. Notice that $\vec{a}_{K}$ does not point towards the origin. The viewing angle is  $\hat{\theta}$ (in this illustration we label angles with the $\hat{}$ ). }
    \label{fig:ill1}
\end{figure}

\begin{figure}
  \centering
    \includegraphics[width=7.truecm]{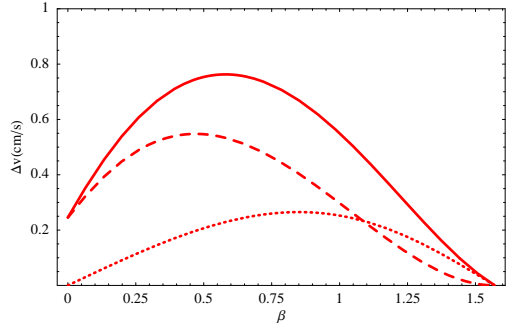}
    \caption{ \small \small Velocity shift curves as a function of $\beta$ for the Kuzmin disc in Newtonian configuration (dashed line), CDM halo (dotted line) and their combination (solid line) for $\theta=0.03$. }
    \label{fig:comb}
\end{figure}

The Kuzmin and the halo accelerations must be projected along the line of sight and then added together. While for spherical symmetric logarithmic potential the acceleration is radial and the angle between the line of sight and $r$ is  simply $\theta'$ (see Fig.~\ref{fig:ill1}), as mentioned before, the projection angle for the  Kuzmin acceleration does not point towards the origin. One clearly has $\cos\theta'\approx \sin\beta$. Using $r\cos\beta\simeq R_g \theta$, $z\simeq\pm\sqrt{r^2-R^2}$ and $R\simeq R_g\theta$, the two line of sight accelerations read 
\begin{eqnarray}
a_{s,K}&=&\frac{MG}{R_{g}^{2}\theta^{2}\Big[1+\Big(|\tan{\beta}|+\frac{h}{R_{g}\theta}\Big)^{2}\Big]}\sin{(\beta\mp\gamma)}\\
a_{s,L}&=&\frac{v_{0}^{2}R_{g}\theta\sqrt{1+\frac{\tan^{2}{\beta}}{q^{2}}}}{R_{c}^{2}+R_{g}^{2}\theta^{2}(1+\frac{\tan^{2}{\beta}}{q^{2}})}\sin{\beta}.
\end{eqnarray}
All the parameters of the gravitational potential will be fixed to the values of \cite{Read:2005}: $M=1.2\cdot 10^{-11}M_\odot$, $h=4.5$ kpc, $v_0=175$ km/s, $R_c=13$ kpc and $q=1$. In Fig.~\ref{fig:comb} the velocity shift caused by the Kuzmin disc and the CDM halo are plotted, as well as their sum, assuming a time span $\Delta T=15$~yr.
Due to its non-radial direction, the main effect of the Kuzmin potential is to shift the maximum and the zero-crossing
of the function with respect to the spherically symmetric case. 
The signal itself is in principle completely independent on the distance of the galaxy; however
we have expressed it as a function of observable quantities, like the viewing angle. In Fig.~\ref{fig:comb} we assume a
distance of the galaxy centre of $800$~kpc, namely the distance of Andromeda galaxy.

\subsection{Disc + MOND}
\label{suboutside}

The Modified Newtonian Dynamics (MOND) paradigm was  first proposed by Milgrom to reconcile discrepancies between general relativity and galaxy scale dynamics \citep{1983ApJ...270..371M}. 
As presented in its  first version, MOND by itself
violates conservation of momentum and energy. In this paper we concentrate on the Bekenstein-Milgrom formulation
of MOND \citep{1984ApJ...286....7B}, which leaves the Newtonian law of motion intact and modifies the standard Poisson equation for the
Newtonian gravitational potential as follows:
\begin{equation}
 \label{mond_pois}
 \nabla \Big[ \mu \Big( \frac{|\nabla \psi|}{a_{0}}\Big)\nabla \psi\Big]=4\pi G \rho,
\end{equation}
where $\psi$  is the MOND gravitational potential, $\nabla^{2}\phi_{N}=4\pi G \rho$, and $a_{0}$ is a scale that was estimated by Milgrom to be $a_{0} = 1.2 \cdot10^{-10}$m/s$^{2}$. The original shape of $\mu$ that helps rendering the right profile of rotational velocities is $\mu(|a|/a_{0}) = 1$ for $|a|\gg |a_0|$ and $\mu(|a|/a_{0}) = |a|/a_{0}$ for $|a| \ll  |a_0|$. 

Eq.~\ref{mond_pois} is a non-linear equation which might be hard to solve analytically, except for a class of symmetric configurations. In \cite{Brada:1995}, a class of disc-galaxy models has been introduced for which exact solutions of the MOND field equation exist.
Subtracting the usual Poisson equation from the MOND equation (\ref{mond_pois}) one has
\begin{equation}
 \label{mond_pois2}
 \nabla \Big[ \mu \Big( \frac{|\nabla \psi|}{a_{0}}\Big)\nabla \psi - \nabla \phi_{N}\Big]=0.
 \end{equation}
For configurations with spherical, cylindrical or plane symmetry the relation between the MOND  field and the Newtonian  field becomes
 \begin{equation}
 \label{mond_pois3}
 \mu \Big( \frac{|\nabla \psi|}{a_{0}}\Big)\nabla \psi = \nabla \phi_{N}.
 \end{equation}
 This equation permits a straightforward relation between the two potentials, such that by assuming a matter density
distribution, one can solve the Poisson equation for the Newtonian potential and then invert it to get the MOND one.

The function $\mu$ defined in MOND has been reformulated during the years, moving from a step function as the
aforementioned type to, e.g., $\mu(x) = x/\sqrt{1+x^{2}}$. As pointed out in \cite{Brada:1995}, the function $I(x) = x\mu(x)$ is therefore
monotonic and invertible, and the inverse is related to the function  $\nu(y) = I^{-1}(y)/y$. Eq.~\ref{mond_pois3} now reads as
\begin{equation}
 \label{out1}
 \nabla \psi=\nu\Big( \frac{|\nabla \phi_{N}|}{a_{0}}\Big)\nabla \phi_{N}.
 \end{equation}

 The exact MOND solution is then given by Eq.~(\ref{out1}), with 
 \begin{equation}
 \label{acc}
 \vec{a}_M=-\nabla\psi=a_{0}I^{-1}\Big(\frac{a_{N}}{a_{0}}\Big)\frac{\vec{a}_{N}}{a_{N}}
   \end{equation}
where $a_M$ and $a_N$ are the MOND and Newtonian acceleration, respectively.

For $\mu(x) = x/\sqrt{1+x^{2}}$, then  $\nu(y) =[1/2 +\sqrt{y^{-2} + 1/4}]$ and consequently
\begin{equation}
 \label{accmond}
 a_{M}=a_{K}\frac{\Big(1+\sqrt{1+\frac{4a_{0}^{2}}{|a_{K}|^{2}}}\Big)^{1/2}}{\sqrt{2}}.
 \end{equation}

The expression for the peculiar acceleration in MOND, outside the disc, is as follows:
\begin{eqnarray}
\label{accmond2}
 a_{s,M}= a_{M}\cdot \sin{(\beta \mp \gamma)},
 \end{eqnarray}
 with
 \begin{eqnarray}
 a_{M}= \frac{MG\Big(1+\sqrt{1+\frac{4a_{0}^{2}}{M^{2}G^{2}}R_{g}^{4}\theta^{4}[1+(|\tan{\beta}|+\frac{h}{R_{g}\theta})^{2}]^{2}}\Big)^{1/2}}{\sqrt{2}R_{g}^{2}\theta^{2}\Big[1+\Big(|\tan{\beta}|+\frac{h}{R_{g}\theta}\Big)\Big]},
 \end{eqnarray}
 and
\begin{equation}
\label{gam}
\gamma=\arccos{\frac{r^{2}+x^{2}-h^{2}}{2 r x}};\qquad x^{2}=r^{2}+h^{2}-2rh\sin{\beta}.
 \end{equation}

\subsection{Comparison}

In order to compare peculiar acceleration predictions between the CDM and the MOND scenarios, it is necessary to match the MOND configuration to a Newtonian one leading to the same rotation velocity curve. 
Given Eq.~\ref{accmond} for the acceleration, the related rotation velocity can be written as \citep{Brada:1995}
\begin{equation}
\label{vel}
v^{2}_{M}(R)=v_{\infty}^{2}\frac{u^{2}}{1+u^{2}}\Big\{ \Big[1+\frac{\zeta^{2}}{4(1+u^{2})^{2}}\Big]^{1/2}+\frac{\zeta}{2(1+u^{2})}\Big\}^{1/2}
 \end{equation}
with $u=R/h$ and  $\zeta=MG/h^{2}a_{0}$ is a measure of how deep in the MOND regime we are. The asymptotic velocity  $v_{\infty}=(MGa_{0})^{1/4}$ is set by the mass of the disc $M=1.2\cdot10^{11} M_{\odot}$  \citep{Read:2005} and has to match the sum of the square velocities of the disc and CDM halo, which are
\begin{equation}
\label{velcdm}
v^{2}_{K}(R)=\frac{GMR^{2}}{(R^{2}+h^{2})^{3/2}},\qquad
v^{2}_{L}(R)=\frac{v_{0}^{2}R^{2}}{R_{c}^{2}+R^{2}},
 \end{equation}
respectively.

\begin{figure}
  \centering
    \includegraphics[width=7.truecm]{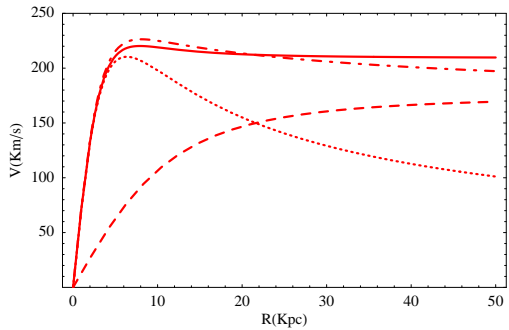}
    \caption{ \small \small Rotation velocity curves for Newtonian configuration (long-short-dashed line) and MOND (solid line). The long-dashed line represents the CDM halo, while the dotted line represents the Kuzmin disc.}
    \label{fig:vel}
\end{figure}

In Fig.~\ref{fig:vel} the rotation curve of the galaxy is drawn both for the Newtonian and MOND configuration. Being the two curves almost undistinguishable, a further interesting point we want to address in this paper is the possibility of discerning a difference between the two gravity models by observing the velocity shift of a test particle outside the disc.
 
\begin{figure}
  \centering
    \includegraphics[width=7.truecm]{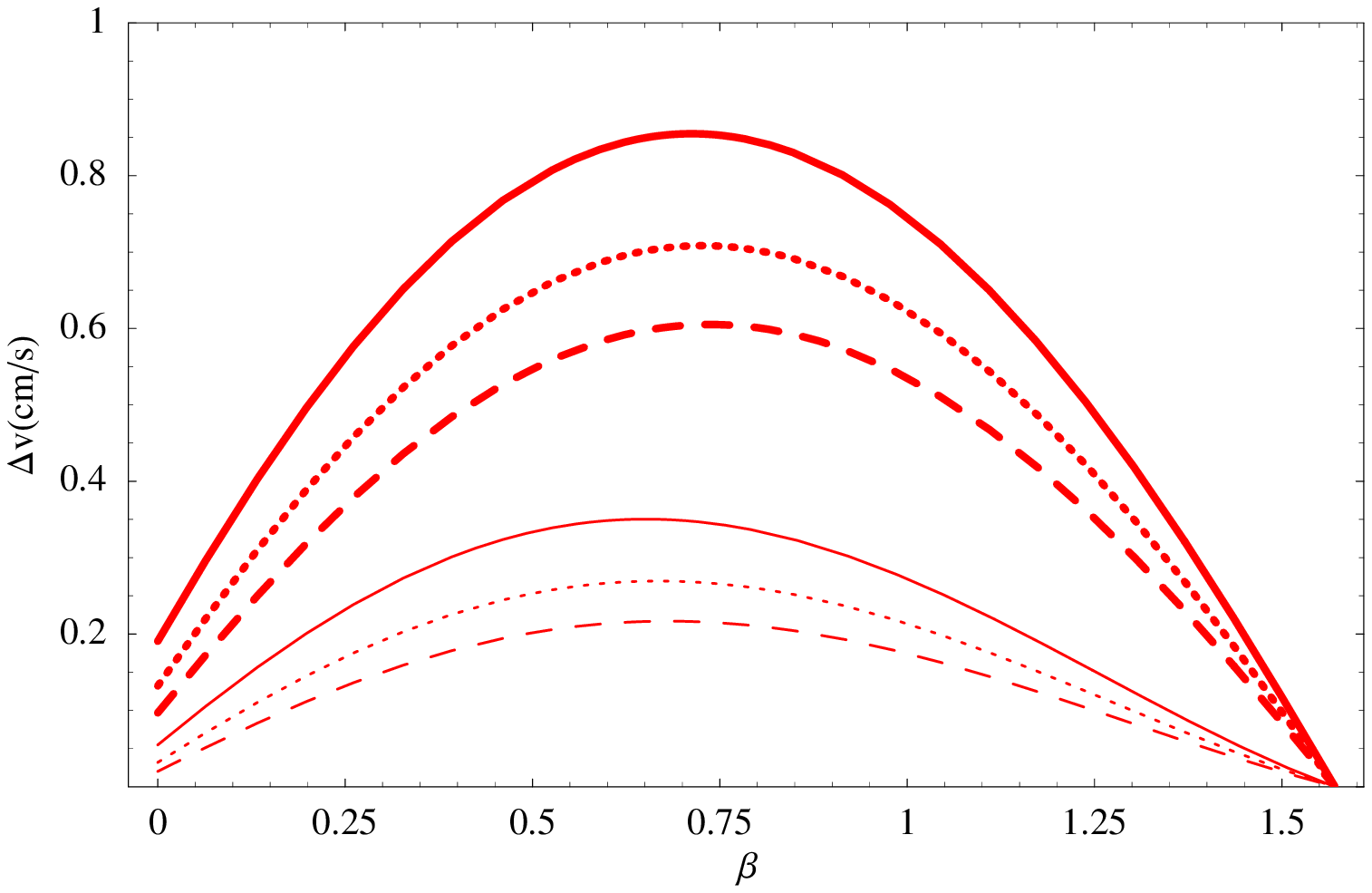}
    \caption{ \small \small  Velocity shift curves caused by peculiar accelerations as a function of $\beta$ for Newtonian configuration (thin lines, CDM halo+Kuzmin disc) and MOND (thick lines). For both configurations curves are drawn for three different values of $\theta$, from top to bottom: $\{0.05,0.06,0.07\}$, corresponding to distances from the centre $\{40,48,56\}$ kpc. }
    \label{fig:pec1}
\end{figure}

\begin{figure}
  \centering
    \includegraphics[width=7.truecm]{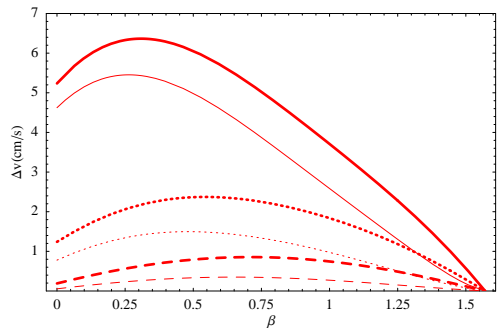}
    \caption{ \small \small Velocity shift curves caused by peculiar accelerations as a function of $\beta$ for Newtonian configuration (thin lines, CDM halo+Kuzmin disc) and MOND (thick lines). For both configurations curves are drawn for three different values of $\theta$, from top to bottom: $\{0.01,0.03,0.05\}$, corresponding to distances from the centre $\{8,24,40\}$ kpc. }
    \label{fig:pec2}
\end{figure}

 The different shape of the peculiar acceleration between Newtonian and MOND configurations are
shown in Fig.~\ref{fig:pec1} and \ref{fig:pec2}. The predicted velocity shift reaches a maximum value along the line of sight,  $\Delta v_{max}$, that depends on the gravitational configuration. The maximum also depends on the distance from the galactic centre, increasing with decreasing distance.  By estimating  $\Delta v_{max}$ from a sample of test particles (e.g.\ globular clusters) outside the disc of the galaxy along each line of sight, one might therefore hope to distinguish the Newtonian and the MOND gravitational potentials.  In our Galaxy, knowing precisely the coordinates of test particles one might trace out the distribution of the signal; in the following section we will focus on this analysis.

\section[]{Milky way}
\label{milkyway}
Now that we have worked out the calculations for the generic spiral galaxy, it is interesting to focus our predictions to the specific case of test particles in the Milky Way. Our Galaxy is known to host almost 150 globular clusters, spread in the stellar halo, with distances up to $100$~kpc, which would constitute an ideal observational target. More importantly, knowing exactly the Galactic coordinates of the objects helps reconstructing and comparing the predicted and the observed peculiar accelerations. 

\begin{figure}
  \centering
    \includegraphics[width=9.truecm]{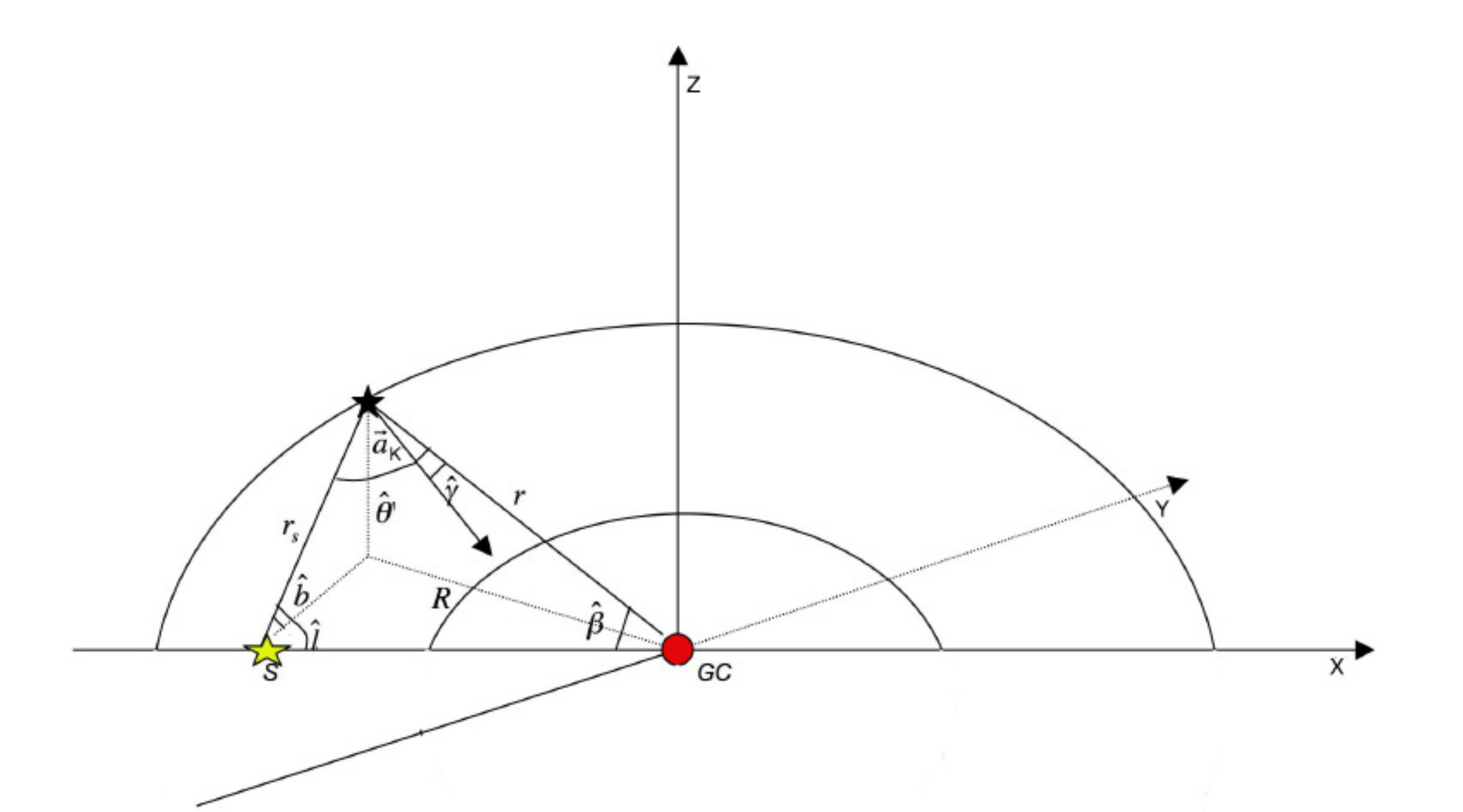}
    \caption{ \small \small Kuzmin disc potential in cartesian coordinates in the Milky Way. The disc lies on plane $(x,y)$. Contours show equal steps of  $\phi_{K}$ and $\vec{a}_K$
represents the radial force  field. The grey star represents the sun and angles are labeled with the $\hat{}$ . }
    \label{fig:ill}
\end{figure}

Let us then write the contribution from peculiar acceleration to velocity shift in terms of Galactic coordinates, in particular the distance of the object from the sun $r_{s}$, the Galactic latitude $b$ and longitude $l$ (see Fig.~\ref{fig:ill}). Firstly, we have to move from cylindrical coordinates $(z, R,\varphi)$ to Galactic coordinates $(r_{s}, b, l)$, and neglecting $\varphi$ due to the axisymmetry of the potential, the transformation reads 
\begin{eqnarray}
\label{transf}
z=r_{s}\sin{b};\qquad R=\sqrt{s^{2}+r_{s}^{2}\cos^{2}{b}-2s r_{s}\cos{b}\cos{l}},
\end{eqnarray}
where $s$ is the distance of the Sun from the Galactic centre.

The line of sight now corresponds to the direction of the distance from the object to us, i.e. $r_{s}$. The projection angle $\theta'$ is related to the new coordinate system as follows:
\begin{equation}
\label{transf}
\cos{\theta'}=\frac{r_{s}^{2}+r^{2}-s^{2}}{2 r_{s}r},
\end{equation}
where $r^{2}=z^{2}+R^{2}$ (see Fig. \ref{fig:ill}).
In addition, as already mentioned in Sec.~\ref{CDM}, for the Kuzmin potential the correction to the non-radial direction of the acceleration must again be taken into account, so that the projection angle changes to $\cos\left({\theta'\pm\gamma}\right)$, and $\gamma$ is as in Eq.~(\ref{gam}), where now
 \begin{equation}
\label{beta}
\beta=\arccos{\frac{r^{2}+s^{2}-r_{s}^{2}}{2rs}}.
\end{equation}

\begin{figure}
  \centering
    \includegraphics[width=9.truecm]{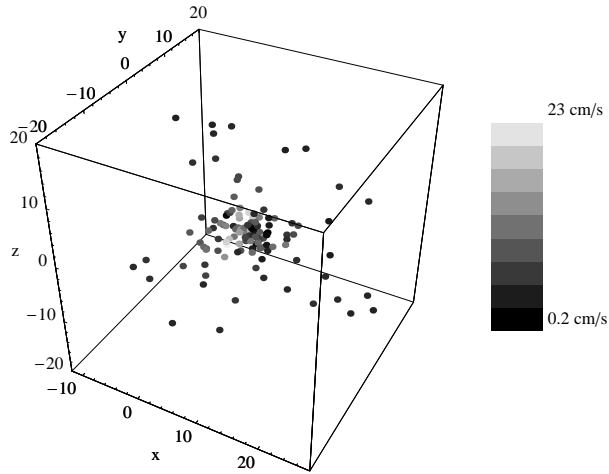}
    \caption{ \small \small The velocity shift signal (for T=15 years) expected in the MOND configuration for the Milky Way globular clusters taken from Harris (1996), plotted in cartesian coordinates (kpc units).}
    \label{fig:mapmond}
\end{figure}

\begin{figure}
  \centering
    \includegraphics[width=9.truecm]{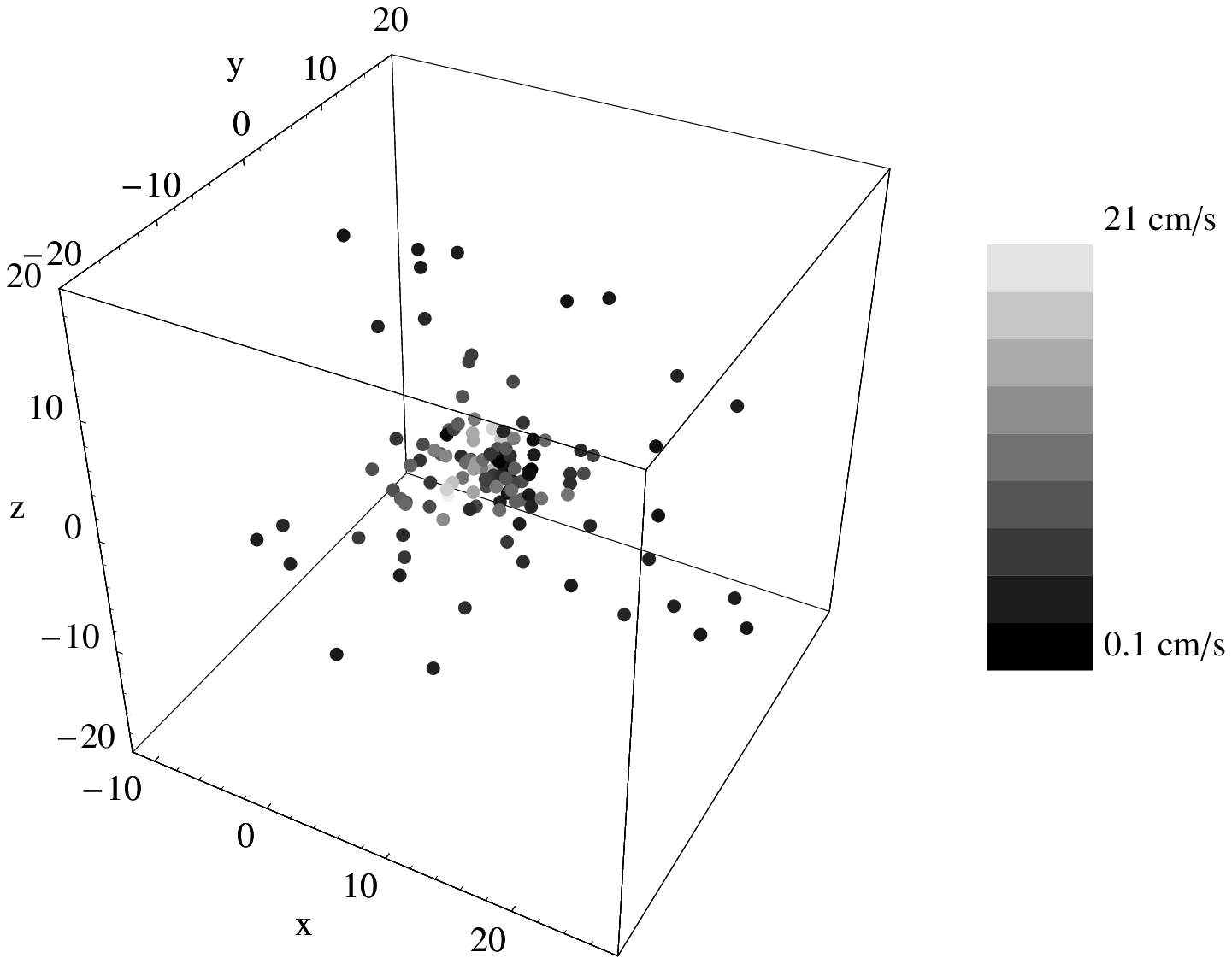}
    \caption{ \small \small The velocity shift signal (for T=15 years)  expected in the CDM halo configuration for the Milky Way globular clusters taken from Harris (1996), plotted in cartesian coordinates (kpc units).}
    \label{fig:mapnewt}
\end{figure}

We have calculated the expected signal for the 150 globular clusters of the \cite{1996AJ....112.1487H} catalogue, excluding only three clusters whose position lies exactly on the disc (where the Kuzmin model breaks down). To be consistent with this catalogue assumption, in our calculations we fix the Sun distance from the Galactic centre to the value $s=8$ kpc.
We map in Fig.~\ref{fig:mapmond} and \ref{fig:mapnewt} the 3D distribution of the velocity shift value for each cluster position, in cartesian coordinates. We find that, over  a time span of 15 years,
 signals as high as 22 cm/s in MOND and 21 cm/s in the Netwonian case can be expected. This is a rather large value when compared to an average cosmological signal of order $5$ cm/s found either for distant galaxies at $z\sim$ 3--4 or for galaxies in clusters (see \cite{2008PhLB..660...81A,2007MNRAS.382.1623B}). We will come back to this result later in Section \ref{conclusions}.
 
\begin{figure}
  \centering
    \includegraphics[width=9.truecm]{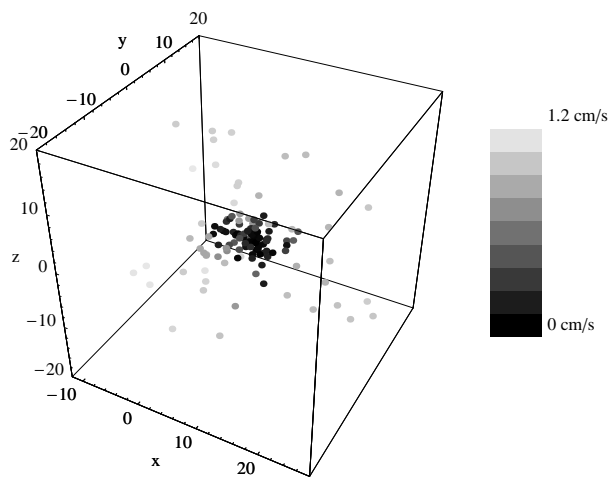}
    \caption{ \small \small The difference in the velocity shift signal (for T=15 years) expected in the CDM halo and MOND configuration for the Milky Way globular clusters taken from Harris (1996), plotted in cartesian coordinates.}
    \label{fig:mapdiff}
\end{figure}

\begin{figure*}
  \centering
    \includegraphics[width=14.truecm]{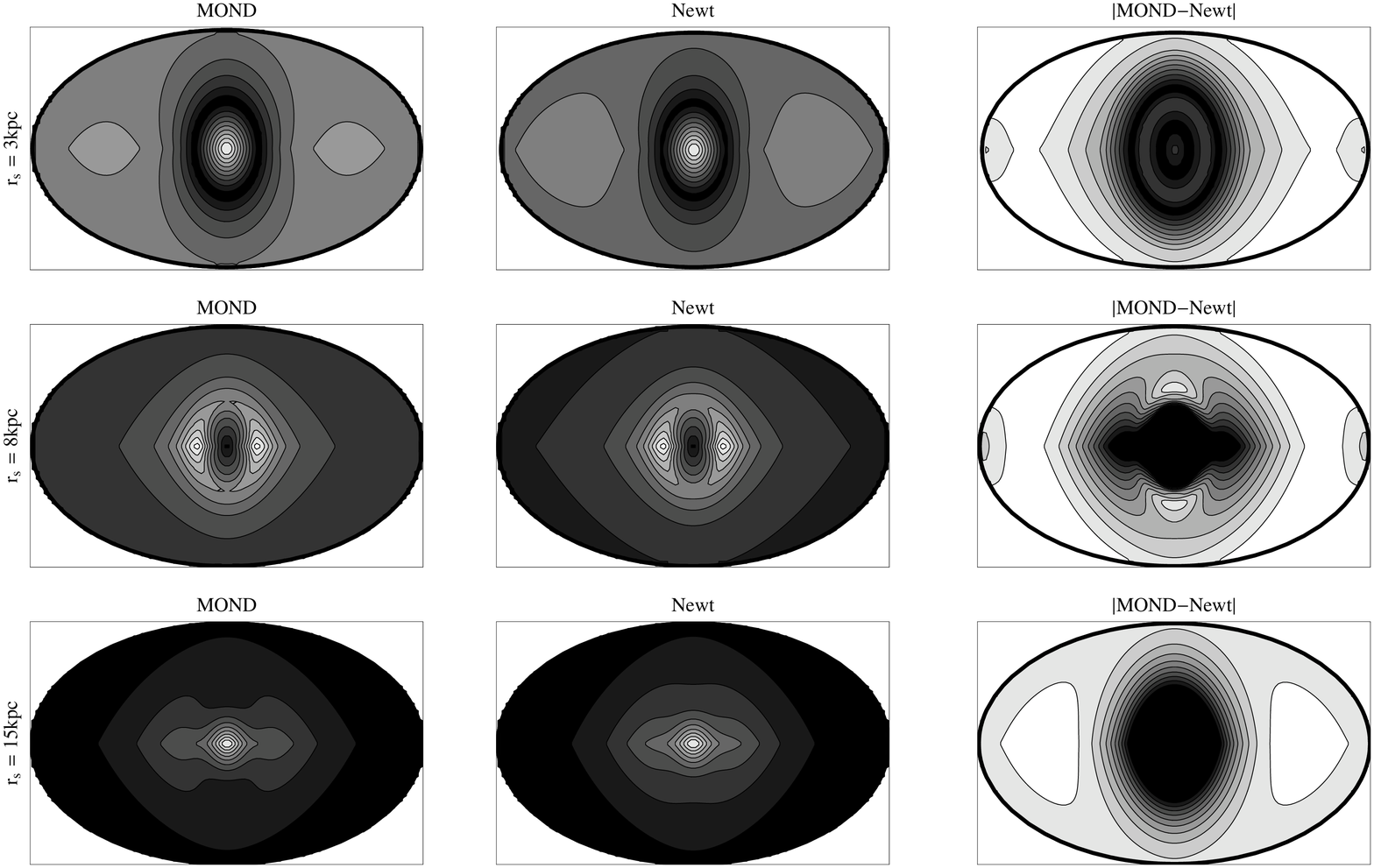}
    \caption{ \small \small Comparison in the velocity shift signal (for T=15 years) as seen in the sky from our position in the Milky Way, for the MOND, CDM halo configuration, and their difference (left to right) and for three distances from the Sun (3, 8 and 15 kpc, top to bottom). The sky signal is contour-plotted  as an equal-area Mollweide projection of the sky around us, where ideal meridians and parallels mark the Galactic longitude and latitude respectively. Regions with higher signal are lighter and the centre of the ellipse corresponds to $\{b,l\}=\{0,0\}$, i.e. to the direction of the Galactic centre.}
    \label{fig:array}
\end{figure*}

In Fig.~\ref{fig:mapdiff} we map the absolute value of the difference between the signals estimated in the two scenarios. Unfortunately, the maximum differences is only of order 1 cm/s, making it quite problematic to observe. However, supposing that such an accuracy will be achieved in the future, it is interesting to note that the two different scenarios result in a distinct morphology of the signal distribution, that could be used as a signature to identify the actual potential. This is shown in Fig.~\ref{fig:array}, for three different distances from the Sun.
 

\section{Discussion and conclusions}
\label{conclusions}

The main purpose of this paper has been to investigate the possibility of reconstructing the gravitational potential of a galaxy by means of the so-called velocity shift signal. Assuming that either the  cosmological redshift of the galaxy is almost vanishing or it can be averaged out, the velocity shift produced by peculiar acceleration of test particles orbiting outside the galactic disc has been the key point of our analysis. In particular we focused on a generic spiral galaxy modelled as a baryonic disc component, e.g.\ the Kuzmin disc, plus a CDM spherical halo. We also extended our calculation to a modified gravity configuration, in which the CDM halo is absent and the Poisson equation is modified following MOND \citep{Brada:1995}: although the two scenarios are undistinguishable using rotation velocity curve, we addressed the question of whether peculiar acceleration might be employed to discriminate between them.

In Sec.~\ref{pec} we have worked out the general treatment for peculiar acceleration in a non-specific galaxy seen by an external observer. The velocity shift signal  has been derived as a function of  the viewing angle $\theta$ and the angle $\beta$ between the radial position and the plane of the disc. The signal reaches a maximum value that depends on the gravitational potential and could in principle be observed along   each line of sight (Fig.~\ref{fig:pec1} and \ref{fig:pec2}, assuming a time span of 15 yr). As expected, the closer is the test particle to the galactic center the higher is the signal, reaching values comparable to the cosmological ones predicted by simulations and possibly detectable from high-resolution and ultra-stable spectrograph coupled to new generation telescope \citep{2008MNRAS.386.1192L}.

In Sec.~\ref{milkyway} we have focused our predictions to our Galaxy and used as test particles the 150  known globular clusters orbiting outside the disc of the Milky Way taken from \cite{1996AJ....112.1487H}. As in this case we clearly are internal observers, the peculiar acceleration has been re-expressed in terms of Galactic coordinates, i.e. the distance from the Sun, the longitude and the latitude. The projected acceleration of the Sun on the disc, which has its maximun value on the order of few cm/s, is assumed to be substracted, and the cluster-centred acceleration of single stars in globular clusters is assumed to be averaged out.   We have mapped out the contours of the predicted signal of both configurations in a Mollweide projection of the sky at three different distances from us (Fig.~\ref{fig:array}). The effect of being off-centred observers  combined with the non-radial direction of the acceleration in Kuzmin disc makes the pattern of the contours non trivial. However, while again the signal reaches the maximum value close to the Galactic centre for both scenarios, the difference signal is stronger at high Galactic longitude (close to $\pi$) where the spherical CDM halo is more influent. This can be better seen by eye in Fig.~\ref{fig:mapmond},\ref{fig:mapnewt} and \ref{fig:mapdiff} where we have mapped the distribution of the peculiar acceleration in Sun-centred cartesian coordinates.

In Table \ref{topten_mond} and  \ref{topten_newt} the first 10 globular clusters exhibiting the highest velocity shift signal are listed. They are the same for CDM halo configuration as for MOND, except the last one.   Their distance is very similar to the distance of the Sun from the Galactic centre, e.g.\ $\sim$8kpc,  and the absolute value of their latitude never exceeds 12$^{o}$: considering also their longitude, it turns out that they are all fairly close to the centre of the Milky Way. As already emphasised, this is not the case for the difference of the signal between the two scenarios. The first 10 globular clusters with the highest difference signal between the CDM halo configuration and MOND are shown in Table \ref{topten_diff}. In view of possible future detections due to the tremendously improved technology \citep{pasquini06,2008MNRAS.386.1192L} this list might indeed give indications on which particular clusters to look at. Although the feasibility of such observations is still under debate, this new technique would open a new window to the ability of testing the gravitational potential and of constraining the parameters involved in its expression, such as the mass of the galaxy and the scale lengths $R_{c}$ and $h$.

Finally, we point out that the advantage of adopting globular cluster as test particles is related  to the fact that they not only are bright, but more importantly composed of thousands of stars and their position is fairly well known; neverthless, in principle one might consider  using other probes, like high velocity clouds or HI regions. Moreover, one could also use the acceleration in the pulsar timing as another probe of the acceleration field in the Galaxy. However this requires further investigations that are beyond the purpose of this paper.

\begin{table*}
 \centering
 \begin{minipage}{140mm}
\caption{The ten globular clusters having the highest velocity shift from peculiar acceleration in the MOND configuration. Astronomical data are from Harris (1996)  and $R_{gc}$ is the distance from the Galactic centre.}
\label{topten_mond}
  \begin{tabular}{@{}llllllll@{}}

\hline
    Name    &  $v$ (cm/s)\footnote{Assuming $T=$15 years} &     RA  (hours)\footnote{J2000 epoch}  & dec (degrees)   &   $l $ (degrees)  &   $b$ (degrees)   &  $r_s$ (kpc) & $R_{gc}$ (kpc)\footnote{Assumes $R_0=8$ kpc} \\
    
\hline

 NGC 6712      &  21.5 &      18 53 04.3  & -08 42 22 &  25.35  & -4.32 &   6.9  & 3.5 \\
 Lynga 7       &20.5      &  16 11 03.0  &-55 18 52  &328.77  & -2.79 &   7.2  & 4.2 \\
 NGC 6535    &18.0       &   18 03 50.7 & -00 17 49 &  27.18 &  10.44  &  6.8  & 3.9 \\
 NGC 5927    &17.9          &15 28 00.5 & -50 40 22 & 326.60 &   4.86  &  7.6 & 4.5 \\
 NGC 6749     &17.3        & 19 05 15.3  & +01 54 03  & 36.20  & -2.20 &   7.9 &  5.0 \\
 NGC 6760    &17.0         & 19 11 12.1  &+01 01 50  & 36.11 &  -3.92   & 7.4 &  4.8\\
 NGC 6352    &16.3        &  17 25 29.2 & -48 25 22&  341.42 &   -7.17 &   5.7 &  3.3 \\
 NGC 6717 (Pal 9)  & 14.9&     18 55 06.2 & -22 42 03  & 12.88 & -10.90   & 7.1  & 2.4 \\
  NGC 6171 (M 107)   &  14.8 &   16 32 31.9  &-13 03 13  &  3.37 &   23.01 &   6.4 &  3.3 \\
 NGC 6541     &     14.0 &    18 08 02.2  &-43 30 00 & 349.48 & -11.09 &   7.0 &  2.2  \\

\hline

\end{tabular}
\end{minipage}
\end{table*}%

\begin{table*}
\centering
 \begin{minipage}{140mm}
\caption{The ten globular clusters having the highest velocity shift from peculiar acceleration in the CDM halo configuration. Astronomical data are from Harris (1996) and $R_{gc}$ is the distance from the Galactic centre.}\label{topten_newt}
  \begin{tabular}{@{}llllllll@{}}

\hline
    Name    &  $v$ (cm/s)\footnote{Assuming $T=$15 years} &    RA  (hours)\footnote{J2000 epoch}  & dec (degrees)   &   $l $ (degrees)  &   $b$ (degrees)   &  $r_s$ (kpc) & $R_{gc}$ (kpc)\footnote{Assumes $R_0=8$ kpc}  \\
    
    \hline

 NGC 6712      &    20.8 &    18 53 04.3  & -08 42 22 &  25.35  & -4.32 &   6.9  & 3.5 \\
 Lynga 7      &19.7      &  16 11 03.0  &-55 18 52  &328.77  & -2.79 &   7.2  & 4.2 \\
 NGC 6535   &17.2        &   18 03 50.7 & -00 17 49 &  27.18 &  10.44  &  6.8  & 3.9 \\
 NGC 5927   & 17.1          &15 28 00.5 & -50 40 22 & 326.60 &   4.86  &  7.6 & 4.5 \\
 NGC 6749  &16.6           & 19 05 15.3  & +01 54 03  & 36.20  & -2.20 &   7.9 &  5.0 \\
 NGC 6760   &16.3          & 19 11 12.1  &+01 01 50  & 36.11 &  -3.92   & 7.4 &  4.8\\
 NGC 6352    &16.0        &  17 25 29.2 & -48 25 22&  341.42 &   -7.17 &   5.7 &  3.3 \\
 NGC 6717 (Pal 9) & 14.5 &      18 55 06.2 & -22 42 03  & 12.88 & -10.90   & 7.1  & 2.4 \\
  NGC 6171 (M 107) &14.0  &     16 32 31.9  &-13 03 13  &  3.37 &   23.01 &   6.4 &  3.3 \\
 IC 1276  (Pal 7)   & 13.7&    18 10 44.2 & -07 12 27 &  21.83 &   5.67  &  5.4  & 3.7 \\
 
 \hline

\end{tabular}
\end{minipage}
\end{table*}%

\begin{table*}
 \centering
 \begin{minipage}{140mm}
\caption{The ten globular clusters having the highest difference in signal between the CDM halo configuration and MOND. Astronomical data are from Harris (1996)  and $R_{gc}$ is the distance from the Galactic centre.}\label{topten_diff}
  \begin{tabular}{@{}llllllll@{}}

\hline
    Name    &   $v$ (cm/s)\footnote{Assuming $T=$15 years} &    RA  (hours)\footnote{J2000 epoch}  & dec (degrees)   &   $l $ (degrees)  &   $b$ (degrees)   &  $r_s$ (kpc) & $R_{gc}$ (kpc)\footnote{Assumes $R_0=8$ kpc} \\
    
    \hline

 Pal 1         &  1.18 &      03 33 23.0  & +79 34 50 & 130.07  & 19.03  & 10.9 & 17.0 \\
 NGC 2298  &1.18       &     06 48 59.2 & -36 00 19 &  245.63 & -16.01 &  10.7 & 15.7 \\
 NGC 1851    &1.17    &      05 14 06.3  & -40 02 50 & 244.51 & -35.04 &  12.1 &  16.7 \\
 NGC 1904 (M 79)  &1.16  &      05 24 10.6 & -24 31 27  & 227.23 &  -29.35  & 12.9  & 18.8 \\
 NGC 288   &1.15   &         00 52 47.5 & -26 35 24 & 152.28 & -89.38 &    8.8 &  12.0  \\
 NGC 5272 (M 3) &1.13   &       13 42 11.2  & +28 22 32 &  42.21 &  78.71 &  10.4  & 12.2 \\
 NGC 5904 (M 5) &1.11  &       15 18 33.8 & +02 04 58  &  3.86  & 46.80 &   7.5  & 6.2 \\
 NGC 1261    &   1.09&       03 12 15.3  & -55 13 01  & 270.54 &  -52.13 &   16.4  & 18.2 \\
 NGC 362   &   1.08&         01 03 14.3 & -70 50 54 & 301.53 &  -46.25  &   8.5  & 9.4 \\
 NGC 7099 (M 30) &1.07   &     21 40 22.0  & -23 10 45 &  27.18 & -46.83 &   8.0  & 7.1\\
 
 \hline

\end{tabular}
\end{minipage}
\end{table*}%

 \bibliography{Bmond}
\end{document}